\documentclass[12pt,amsfonts]{article}    

\usepackage{amsfonts}

\def\ts{\textstyle}
\def\vp{\varphi}
\def\bl{\mathbf{\l}}

\def\half{\textstyle{\frac{1}{2}}}

\def\H{{\cal H}}

\def\mfH{\mathfrak{H}}

\def\p{\varphi}

\def\H{{\cal H}}

\def\l{\lambda}

\def\t{\textstyle}

\def\ra{\rightarrow}
\def\tint{{\textstyle\int}}
\def\hg{{\hat g}}
\def\hp{{\hat\pi}}
\def\hph{{\hat\varphi}}
\def\s{\hskip.08em}
\def\d{\partial}

\def\b{\begin{eqnarray*}}  
\def\e{\end{eqnarray*}}    
\def\bn{\begin{eqnarray}}  
\def\en{\end{eqnarray}}   

\def\<{\langle}
\def\>{\rangle}

\def\bk{\mathbf k}
\def\bm{\mathbf m}

\def\de{\delta}
\def\no{\nonumber}

\def\ds{d^s\!x}
\def\k{\kappa}
\def\bl{\mathbf l}
\def\quarter{\textstyle{\frac{1}{4}}}

\def\hk{\hat{\kappa}}
\def\{{\lbrace}
\def\hv{\hat{\varphi}}
\def\}{\rbrace}
\begin{document}

\title{Using Affine Quantization to Analyze \\Non-renormalizable Scalar Fields and\\
the Quantization of Einstein's Gravity}          
\author{John R. Klauder\footnote{klauder@
phys.ufl.edu} \\
Department of Physics and Department of Mathematics \\
University of Florida,   
Gainesville, FL 32611-8440}
\date{ }
\bibliographystyle{unsrt}

\maketitle 
\begin{abstract}
Affine quantization is a parallel procedure to canonical quantiza- tion, which is ideally suited to deal with non-renormalizable scalar models as well as quantum gravity. The basic applications of this approach lead to the common goals of any quantization, such as Schr\"odinger's representation and Schr\"odinger's equation. Careful at- tention is paid toward seeking favored classical variables, which are those that should be promoted to the principal quantum operators. This effort leads toward classical variables that have a constant pos- itive, zero, or negative curvature, which typically characterize such favored variables. This focus leans heavily toward affine variables with a constant negative curvature, which leads to a surprisingly ac- commodating analysis of non-renormalizable scalar models as well as Einstein's general relativity.
\end{abstract}

\section{Introduction}
\subsection{A brief look at three quantization procedures}
Canonical quantization is traditionally used to quantize most classical theories. For a simple system, a favored pair of phase-space variables, i.e., $p$ and $q$, for which $-\infty<p,q<\infty$, and which are also Cartesian coordinates, arising from a flat surface \cite{dirac}, i.e., a
 constant zero curvature surface, 
to become $P$ and $Q$, the basic pair of quantum variables, with $[Q,P]=i\hbar 1\!\!1$.

Another familiar approach deals with the $SU(2)$ or $SO(3)$ groups, and its favored classical variable pair arises from a spherical surface, i.e., a constant positive curvature surface of fixed radius determined by the Hilbert space dimension \cite{j1}, along with its basic operators $S_1,S_2$, and $S_3$, such that $[S_1,S_2]=i\hbar S_3$ with valid permutations.

A third example, which is less well known, involves affine quantization,
that, in one example, involves a favored pair of phase-space variables, $p$ and $q$, for $-\infty<p<\infty$ while $0<q<\infty$, and the geometric surface is that of a constant negative curvature \cite{cnc}, along with the basic pair of operators $0<Q<\infty$ and $D=(QP+PQ)/2$, which
fulfills $[Q,D]=i\hbar Q$ \cite{j1,j2}.\footnote{The affine variable $Q$ can instead satisfy
$-\infty<Q<0$ or even $-\infty<Q\neq0 <\infty$, a reducible operator that the program of enhanced quantization
permits \cite{eq}. The word `affine' has been chosen for the 
similarity to an affine group, especially the  symbolic equality of their Lie algebras \cite{22}.}
\subsubsection{Favored classical variables}

Favored phase-space coordinates to promote to quantum operators apply to all three quantization procedures. To illustrate the meaning of favored coordinates we examine an example from canonical quantization. A classical harmonic oscillator Hamiltonian, say $H(p,q)=(p^2+q^2)/2$, in one set of coordinates,
can also be described by alternative phase-space coordinates, say $\tilde{p}$ and $\tilde{q}$, as one example, where $p=\tilde{p}/\tilde{q}^2$ and
$q=\tilde{q}^3/3$. It follows that $H(p,q)=\tilde{H}(\tilde{p},\tilde{q})=(\tilde{p}^2/\tilde{q}^4+
\tilde{q}^6/9)/2$. Although the quantum operators obey $[Q,P]=i\hbar1\!\!1=[\tilde{Q},\tilde{P}]$, it follows that  $\hat{H}(P,Q)=(P^2+Q^2)/2\neq \hat{\tilde{H}}(
\tilde{P},\tilde{Q})=(\tilde{P}^2/\tilde{Q}^4+\tilde{Q}^6/9)/2$.  The spectrum of these two 
Hamiltonians are different despite the fact that they agree when $\hbar\ra0$; here, apart from linear transformations, one choice of phase-space variables is correct, while any other choice of phase-space variables is incorrect, and that difference may already show up at the lowest order of $\hbar\neq0$.

It is essential to identify the favored classical variables, and only promote them to quantum operators; otherwise you risk a false quantization! We now focus on affine quantization.

\subsection{The essence of affine quantization}
Canonical quantization is the standard approach, but it can fail to yield an acceptable quantization, such as for a classical `harmonic oscillator' with $0<q<\infty$. This very problem is easy to quantize with affine quantization; see \cite{j1}. Coherent states for affine quantization, with positive $q$ and $Q$ having passed their dimensions to $p$ (or carried by $D$), rendering them dimensionless for simplicity, are given by
     \bn |p;q\>\equiv e^{ipQ/\hbar}\,e^{-i\ln(q)D/\hbar}\,|b\>\;,  \en
with $[(Q-1)+iD/b\hbar]\,|b\>=0$. If ${\cal{H}}'(D,Q)$ denotes the quantum Hamiltonian, then,
a semiclassical expression called the  `weak correspondence principle' \cite{ww} is given by
     \bn &&H(p,q)\equiv H'(pq,q)=\<p;q| {\cal{H}}'(D,Q)|p;q\>  \no \\
          &&\hskip7.6em=\<b|{\cal{H}}'(D+pqQ,qQ)|b\> \\
          && \hskip7.6em={\cal{H}}'(pq,q)+{\cal{O}}(\hbar;p,q)\;, \no \en
implying that when $\hbar\rightarrow 0$, leading to the standard classical limit,
then $H'(pq,q)={\cal{H}}'(pq,q)$; namely, the
quantum variables have the same functional positions as the appropriate classical variables. 
In addition, we find that these variables lead to a constant negative curvature surface (equal to $-2/b\hbar$) as shown by the equation\footnote{Similar stories for canonical and spin quantizations appear in \cite{j1}.}
   \bn  d\sigma(p,q)^2\equiv 2\hbar[|\!|\, d|p;q\>|\!|^2 -|\<p;q|\,d|p;q\>|^2]=(b\hbar)^{-1}q^2\,dp^2+(b\hbar)\, q^{-2}\,dq^2 \;. \en
   This latter property, i.e., seeing that these particular classical variables arise from a constant negative curvature renders them as favored coordinates, just like the favored variables of canonical quantization are those that are Cartesian coordinates, i.e., a constant zero curvature \cite{dirac}.

 After this background, we turn attention to the Schr\"odinger representation and equations for affine quantization. The quantum action functional ($q$), with normalized Hilbert space vectors, is given by
  \bn A_q=\tint_0^T \<\Psi(t)|[i\hbar(\d/\d t)-{\cal{H}}'(D,Q)]|\Psi(t)\>\;dt  \;, \en
  and variational efforts lead to a form of Schr\"odinger's equation
    \bn i\hbar\,(\d |\Psi(t)\>/\d t)={\cal{H}}'(D,Q)\,|\Psi(t)\> \;. \en
    Schr\"odinger's representation is $Q\rightarrow x$ and
    $D\rightarrow -i\half\hbar[x(\d/\d x)+(\d/\d x)x]=-i\hbar[x(\d/\d x)+1/2]$,
    where $0<x<\infty$ (provided $0<Q<\infty$), and $|\Psi(t)\>\rightarrow \psi(x,t)$.
    This analysis leads to the familiar form of the Schr\"odinger equation
    \bn i\hbar\,\d\,\psi(x,t)/\d t={\cal{H}}'(-i\hbar[x(\d/\d x)+1/2],x)\,\psi(x,t)\;.\en
    
    There is a new feature in affine quantization, one that is not in canonical quantization, namely that 
    \bn Dx^{-1/2} =-i\hbar[x(\d/\d x)+1/2]\,x^{-1/2}=0 \:. \label{444} \en
    The analog of this relation in canonical quantization is $P1\!\!1=-i\hbar(\d /\d x)\,1\!\!1=0$, which is self-evident, and leads to no useful relation.
    
    The equations above, dealing with some basic properties, have their analogues
    in more complex systems, which are analyzed next in Sec.~2 regarding quantizing non-renormalizable scalar fields, followed by Sec.~3 regarding quantizing gravity.   
    
\section{Canonical and Affine Quantization of \\ Non-renormalizable Scalar Fields}
\subsection{Possible results from canonical quantization}
The conventional version of covariant scalar fields deals with the quantization of models given by
the classical Hamiltonian
 \bn  H_c(\pi,\varphi)=\tint\{ \half[\pi(x)^2+({\overrightarrow{\nabla}}\varphi)(x)^2+
 m_0^2\,\varphi(x)^2\,]+g_0\,\varphi(x)^p\,\}\;d^s\!x\;, \label{ss} \en
where $p$ is the (even positive integer) power of the interaction term, $s$ is the (positive integer) number of spatial dimensions (with 
$n\equiv s+1$ as the number of spacetime dimensions), $m_0^2>0$ is the mass term, and $g_0\geq0$ is
the coupling constant. 

Canonical quantization leads to expected results for `free models' (i.e., $g_0=0$) and all $n\geq2$,
while  `non-free models' (i.e.,  $g_0>0$) require that $p<2n/(n-2)$. The case of $p=4=n$ was determined to `become free' by Monte Carlo studies 
\cite{82}, which probably would also apply to the case $p=6$ and $n=3$. The
remaining models, where $p>2n/(n-2)$,  are non-reormalizable and, following a perturbation expansion of $g_0$ there is an
infinite number of different, divergent terms;  or, if treated as a whole, such models collapse to 
`free theories' with a vanishing interaction term \cite{AA,FF}. 

Briefly summarized, canonical quantization leads to unacceptable results whenever $p>2n/(n-2)$. On the other hand, a classical analysis of cases where $p>2n/(n-2)$ leads to natural and expected results. 

We now show how models for which $p>2n/(n-2)$ can be successfully quantized using affine quantization rather than canonical quantization.

\subsection{Possible results from affine quantization} 
The classical Hamiltonian in (\ref{ss}) is the same starting point, except that we 
require that $\vp(x)\neq0$ and replace the momentum field $\pi(x)$ with the affine field $\kappa(x)\equiv \pi(x)\,\varphi(x)$, which leads to the affine version of the classical Hamiltonian given by
   \bn H'_c(\kappa,\varphi)=\tint \{\half[ \kappa^2(x)\varphi(x)^{-2}+({\overrightarrow{\nabla}}\varphi)(x)^2+ m_0^2\,\varphi(x)^2\,]+g_0\,\varphi(x)^p\,\}\;d^s\!x\;, \label{kl} \en
   and the parameters $p$, $s$, $m_0^2>0$, and $g_0\geq0$ have the same meaning as before. The 
   Poisson bracket $\{\varphi(x),\kappa(x')\}=\delta^s(x-x')\,\varphi(x)$, with $\varphi(x)\neq0$ (see footnote 1), points toward the
   commutator $[\hat{\varphi}(x),\hat{\kappa}(x')]=i\hbar\,\delta^s(x-x')\,\hat{\varphi}(x)$,
   with $\hat{\varphi}(x)\neq0$. 
   
   The Schr\"odinger representation is $\hat{\varphi}(x)=\varphi(x)\neq0$ and 
    \bn \hat{\kappa}(x)=-i\half\hbar[ \varphi(x)(\delta/\delta \varphi(x))+(\delta/\delta\varphi(x))\varphi(x)]\;,\en
   which leads to an affine Schr\"odinger quantization of the classical affine Hamiltonian given by
    \bn &&{\cal{H}}'(\hk, \hv)=\tint \{ \half[ \hat{\kappa}(x)\varphi(x)^{-2}
    \hat{\kappa}(x)+ ({\overrightarrow{\nabla}}\varphi)(x)^2+ m_0^2\,\varphi(x)^2\,] \no \\
    &&\hskip14em+g_0\,\varphi(x)^p\,\}\;d^s\!x\;,  \label{gh} \en
   and which appears to be only a `formal representation and equation', since it is true that
   $\delta\varphi(x')/\delta\varphi(x)=\delta^s(x'-x)$, leads to $\infty$ if $x'=x$.

   These functional derivatives are derived from regularized procedures which replace $\varphi(x)$ 
   with a discrete basis that treats all of $x$ as an $s$-dimensional lattice so 
   $\varphi(x)\rightarrow \varphi_{\bf k}$, and the normal space $x \rightarrow {\bf k}a$,
    ${\bf k}\in \{\cdots, -1, 0, 1, 2, 3, \cdots\}^s$, and $a>0$ is the physical distance between rungs of the lattice. In this regularization, 
  \bn \hk_\bk=-i\s\half\s\hbar\s[\s\p_\bk\s(\d/\d\p_\bk)+ (\d/\d\p_\bk)\s\p_\bk\s]\s\s a^{-s}\;.
  \label{rrr} \en 
  Additionally, $a^s$ is a tiny physical volume, and $ba^s$ (with $b\simeq 1$) is a tiny dimensionless volume. This expression leads to $\hk_\bk\,\p_\bk^{-1/2}=0$,
    which, in the limit  $a\ra0$, leads to ${\kappa}(x)\,\varphi(x)^{-1/2}=0$ 
    (see (\ref{444})). 
    
    This analysis points toward a regularized ($r$) quantum Hamiltonian given by
    \bn  &&\hskip-1.1em\H_{r}=\half\s{\t\sum}_\bk\,\hk_\bk\,(\p_\bk^2)^{-(1-2ba^s)}\,\hk_\bk\,a^s+\half{\t\sum}_{\bk,\bk^*}(\p_{\bk^*}-\p_\bk)^2\s a^{s-2}   \no\\
   &&\hskip4em +\half\s m_0^2{\t\sum}_\bk\p_\bk^2\,a^s+
    g_0{\t\sum}_\bk\p_\bk^p\,a^s-E_0\;, \label{eH2}\en
    where $\bk^*$ is one positive step forward from the site $\bk$ for each of the $s$ nearest lattice sites, in which the site labels may be  spatially  periodic. Equation (\ref{eH2}) is the
    first example of a regularized Hamiltonian. 
    
    A second example of a regularized Hamiltonian
    is given, with  $J_{\bk,\bl}\equiv 1/(2s+1)$ for $\bl =\bk$ and the $2s$ nearest spacial neighbors to $\bk$, by 
  \bn  &&\hskip-1.1em\H'_{r}=\half\s{\t\sum}_\bk\,\hk_\bk\,
   (\Sigma_\bl J_{\bk,\bl}\p_\bl^2)^{-(1-2ba^s)}\,
   \hk_\bk\,a^s+\half{\t\sum}_{\bk,\bk^*}(\p_{\bk^*}-\p_\bk)^2\s a^{s-2}   \no\\
   &&\hskip4em +\half\s m_0^2{\t\sum}_\bk\p_\bk^2\,a^s+
    g_0{\t\sum}_\bk\p_\bk^p\,a^s-E'_0\;. \en
    
    A different kind of regularization offers a third regularized Hamiltonian operator given by
      \bn &&\hskip-1.1em\H''_r=-\half\hbar^2\s a^{-2s}{\t\sum}_\bk\,\frac{\d^2}{\d\p_\bk^2}\,a^s+\half{\t\sum}_{\bk,\bk^*}(\p_{\bk^*}-\p_\bk)^2\s a^{s-2} \no \\
    &&\hskip3em+\half\s m_0^2{\t\sum}_\bk\p_\bk^2\,a^s+
    g_0{\t\sum}_\bk\p_\bk^p\,a^s
    +\half\hbar^2{\t\sum}_\bk\s {\cal F}_\bk(\p)\,a^s-E_0\;. \label{eH}\en
    In this expression, the counterterm is proportional to $\hbar^2$, and specifically is chosen so that
    \bn   &&\hskip-2.9em {\cal F}_\bk(\p) \equiv \frac{a^{-2s}}{{\Pi_\bl}[\Sigma_{\bm}\s J_{\bl,\bm}\s \varphi_\bm^2]^{-(1-2ba^s)/4}}\,
    \frac{\d^2\,\Pi_\bl\s[\Sigma_{\bm}\s J_{\bl,\bm}\s \varphi_\bm^2]^{-(1-2ba^s)/4}} {\d\s\varphi^2_\bk}\no \\
    &&=\quarter\s(1-2ba^s)^2\s
          a^{-2s}\s\bigg({{\ts\sum}_{\s \bl}}\s\frac{\t
  J_{\bl,\s \bk}\s \p_\bk}{\t[\Sigma_\bm\s
  J_{\bl,\s \bm}\s\p_\bm^2]}\bigg)^2\no\\
  &&\hskip1em-\half\s(1-2ba^s)
  \s a^{-2s}\s{{\ts\sum}_{\s \bl}}\s\frac{\t J_{\bl,\s \bk}}{\t[\Sigma_\bm\s
  J_{\bl,\s \bm}\s\p^2_\bm]} \no\\
  &&\hskip1em+(1-2ba^s)
  \s a^{-2s}\s{{\ts\sum}_{\s \bl}}\s\frac{\t J_{\bl,\s \bk}^2\s\p_\bk^2}{\t[\Sigma_\bm\s
  J_{\bl,\s \bm}\s\p^2_\bm]^2}\;. \label{eF} \en
  
  \subsection{Affine coherent states for covariant scalar fields}
  In choosing suitable coherent states we need to deal with the fact that 
   $-\infty<\vp(x)\neq0<\infty$ as well as $-\infty<\hv(x)\neq0<\infty$,
   where 
   \bn \Pi_x[(\hv(x)-1)+i\hk(x)/\beta\hbar]\,|\beta\>=0 \;.\en
   The coherent states then become
     \bn |\pi;\vp\>=e^{(i/\hbar)\tint \pi(x)\,\hv(x)\,\ds}\;e^{-(i/\hbar)\tint \ln(|\vp(x)|)
     \,\hk(x)\,\ds}\;|\beta\> \;, \en
     and the semiclassical Hamiltonian is given by
     \bn &&\hskip-9em H(\pi,\vp)=\<\pi;\vp|{\cal{H}}(\hk,\hv)
     |\pi;\vp\> \no \\
     &&\hskip-5em=\<\beta|{\cal{H}}(\hk(x)+\pi(x)|\vp(x)|\hv(x),|\vp(x)|\hv(x))|\beta\> \no \\                  
     &&\hskip-5em =\<\beta|{\cal{H}}(\hk(x)+\pi(x)\vp(x)|\hv(x)|,\vp(x)|\hv(x)|)|\beta\> \no \\
     &&\hskip-5em={\cal{H}}(\pi,\vp)+{\cal{O}}(\hbar; \pi, \vp) \;. \en
     
     For a suitable $L$ it follows that
     \bn &&\hskip-6em
     d\sigma(\pi,\vp)^2=L\hbar[\,|\!|\,d|\pi;\vp\>|\!|^2-|\<\pi;\vp|\,d|\pi,;\vp\>|^2\,]\no \\
     &&\hskip-1.6em =\tint\{(\beta\hbar)^{-1}\,\vp(x)^2\,d\pi(x)^2+(\beta\hbar)\,\vp(x)^{-2}\,d\vp(x)^2\,\}\;\ds
     \:. \en
     The result is a constant negative curvature,
     namely $-2/\beta\hbar$, for each and every point $x$.
     
     \subsection{Arguments supporting non-renormalizable behavior}
     An important feature of many 
     non-renormalizable models is the fact that reducing
      the intersection term to zero does not return the model to a free theory. This unusual feature can be illustrated on a toy model
      the basic Hamiltonian of which is given for $-\infty< p,q<\infty$ and $g_0\geq$ by
  \bn H(p,q)=\half(p^2+q^2)+g_0q^{-4}\;, \en
  which, if $g_0=0$ appears to be a free harmonic  oscillator. However, that is deceptive because if that $g_0$ is turned on, i.e., $g_0>0$, and then turned off, namely $g_0\ra0$, it follows 
  from continuity that $q=0$ is forbidden, namely $-\infty<p<\infty$ but now $-\infty<q\neq0<\infty$;
   the result can be called a `pseudofree theory'. That may seem to be
   a tiny change, but the spectrum of the free and the pseudofree quantum theories becomes markedly different. Instead of
   the free ($f$) theory propagator, which is given by
   \bn K_f(q'',T;q',0)=\sum_{n=0,1,2,3,\cdots} h_n(q'')\,h_n(q')\,e^{-i(n+1/2)T/\hbar} \;,\en
    where $h_n(q)$ are the Hermite functions, the pseudofree ($pf$) theory propagator is instead given by
    \bn K_{pf}(q'',T;q',0)=2\,\theta(q''q') \sum_{n=1,3,5,7,\cdots}h_n(q'')h_n(q')\,e^{-i(n+1/2)T/\hbar}
    \;, \en
    with $\theta(u)=1$ if $u>0$, while $\theta(u)=0$ if $u<0$. Clearly, a perturbation about the free theory leads to unlimited divergences, while a perturbation about the given pseudofree theory leads to an acceptable approach to study this example. The lesson that this toy model offers is that {\it 
    domains matter};
    the domain here being the set of continuous functions, $\{p(t),q(t)\}_0^T$, $T>0$, for which $\tint_0^T H(p(t),q(t))\,dt<\infty$.
    
     A different example also demonstrates that the quantum theory of a nonrenormalizable model is
   connected to a pseudofree quantum version and not to its free quantum version. The model in question is that of an ultralocal ($u$) scalar field, and its affine classical Hamiltonian is given by
   \bn H'_u(\k,\p)=\tint\{\half[\k(x)^2\vp(x)^{-2}+m_0^2\,\p(x)^2] +g_0\,\p(x)^p\,\}\;d^s\!x\;, \en
   which differs from (\ref{ss}) because the gradient term is gone. Clearly, for every example 
   with $p>2$, the domain for the interacting version is smaller than the domain for the 
   non-interacting version.

   The Schr\"odinger representation involves $\hph(x)\ra \vp(x)\neq0$ and
    \bn \hk(x)\ra -i\half\hbar[\vp(x)(\de/\de\vp(x))+(\de/\de\vp(x))\vp(x)]\;. \en
    Then the regularized quantum Hamiltonian for this model is given by
    \bn {\cal{H}}'(\hk,\hph)=\sum_\bk      \{\half[\hk_\bk\,(\vp_\bk)^{-2}\hk_\bk+m_0^2\,\vp_\bk]+g_0\,
    \vp_\bk^p\,\}\;a^s\;. \en
    With (\ref{rrr}) as $\hk_\bk$, then  $\hk_\bk\,\p_\bk^{-1/2}=0$. Schr\"odinger's equation,  $i\hbar \d \,                                 
    \psi(\p,t)/\d t={\cal{H}}'(\hk,\hv)\,\psi(\p,t)$,
    and the regularized ground state is given by
       \bn \psi_0(\vp)={e^{-W(\vp)/2}}\,\Pi_\bk [(ba^s)^{-1/2}|\vp_\bk|^{-(1-2ba^s)/2} ]\;, \en
       where $W(\vp)$ is real.
       
       The characteristic function, i.e., the Fourier transform of the normalized 
       version of $|\psi_0(\vp)|^2$
       for this model, takes the form
\bn  &&C(f)=\lim_{a\ra0}\s\Pi_\bk\s\tint \,\{ e^{ i f_\bk \vp_\bk/\hbar}\,e^{-W(\vp_\bk)}\, ( ba^s)\,|\vp_\bk|^{-(1-2ba^s)}\s\}\, d\vp_\bk \no\\
     &&\hskip2.48em   =\lim_{a\ra0}\s\Pi_\bk\,\{1 - (ba^s)\tint [1- e^{ i f_\bk \vp_\bk/\hbar} ]\,\,e^{-W(\vp_\bk)} \,|\vp_\bk|^{-(1-2ba^s)\,}d\vp_\bk \}\no\\  &&\hskip2.48em =\exp\{-b\tint d^s\!x\s\tint[1-e^{if(x)\s\l/\hbar}] e^{-w(\l,\hbar)}\,d\l/|\l|\}\;, \label{tyu} \en
     where $\vp_\bk\ra\l$, and $w$ may involve parameter renormalization as well. The result in 
     (\ref{tyu}), which, besides a Gaussian distribution, is the only other outcome of the Central Limit Theorem, and is called a
     (generalized) Poisson distribution. For this solution, as $g_0\ra0$, the factor
     $w(\l,\hbar)\ra c\l^2$, where $c>0$, which leads to the pseudofree solution for this example.
     
     The example of a field theory without any gradients has led to a well-defined continuum result. This result points to reasonable continuum limits for the earlier models that do have gradients, which will even soften the analysis.
     
     \subsection{Computer studies of  non-renormalizable models}
     A Monte Carlo (MC) study, by Freedman, Smolensky, and Weingarten in 1982 \cite{82}, 
     examined two covariant scalar fields of the $\vp^p_n$ type, where $n$ is the spacetime dimension. This study for $p,n=4,3$ confirmed 
     a proper quantization of that scalar field, and, as well, showed that a proper quantization of a
     $p,n=4,4$ model failed and instead that it led to a free theory. A MC study of a $p,n=4,4$ model using the regularized version shown in (\ref{eH}) and (\ref{eF}) has given a hint that such a 
     regularization may offer a positive result. However, such studies can take considerable time and effort. A less time-consuming model of a conventional non-renormalizable  model, namely $p,n=8,3$
     has begun but not yet points to whether or not the same regularized version would be a success or a falure in overcoming its conventional non-renormalizability.
     
     The author of this paper urges
     additional MC studies by others to see if any of the proposed regularized versions of non-renormalizable models presented in this paper could lead to acceptable quantizations.

     \section{Canonical and Affine Quantization of \\ Einstein's Gravity}
     \subsection{Canonical quantization and Einstein's gravity}
     Classical general relativity, as defined by Einstein, is a marvelous theory that has 
     proven to be correct in a variety of ways. The standard phase-space variables \cite{adm}, namely the spacial metric field $g_{ab}(x)$ (symmetric in $ab$) and the spacial momentum 
     field $\pi^{cd}(x)$ (symmetric in $cd$), and where
     $a,b,c,d =1,2,3$, prove difficult to quantize since the classical metric is 
     strictly positive, e.g.,  $ds^2=g_{ab}(x)\,dx^a\,dx^b>0$. Using canonical quantization 
     is limited to a successful result only if all the classical variables can assume arbitrary
     values between $-\infty$ and $+\infty$. Efforts to get around these difficulties have led
     to deviations from the original general relativity by adding higher powers of the scalar curvature, adding additional derivatives to the equations of motion, 
     non-commuting spacial variables, as well as factorizing the 
     metric field into the product of two terms, i.e., $g_{ab}(x)=E_a^i(x)\delta_{ij}E_b^j(x)$, with
     $i,j=1,2,3$,
     and where  $\delta_{ij}$ is $1$ if $i=j$, or is $0$ if $i\neq j$; these variables also appear 
     with modest variations. In this case $E_a^i(x)$ obeys 
     the rule to be between $-\infty$ and $+\infty$;  but, these  rules also allow some $E_a^i(x)=0$, 
     in which case the metric $g_{ab}(x)\,dx^a\,dx^b\geq0$, and fails to be strictly positive.
      Moreover, choosing (a slight variation of) $E_a^i(x)$ and a natural partner $A_i^a(x)$, which have a constant
      for their Poisson bracket, become candidate partners to promote to the basic pair of quantum operators. If these two classical variables were also suitable to be Cartesian coordinates, as Dirac has observed \cite{dirac}, then they could be favored variables. Unfortunately, the variables $E_a^i(x)$ and $A^a_i(x)$, which are primary variables in the program of `loop quantum gravity'
      (see, e.g., \cite{aa, ghj, 111, 123}),
are not suited to be a pair of Cartesian coordinates, which then implies that quantization of these two variables would lead to a false quantization \cite{jk4}. 

Moreover, in several ways, loop quantum gravity is different than traditional  (i.e., canonical or affine) quantization. This is because the loops and their intersection are important and play a significant role, space is also discrete, etc. 

On the other hand, affine quantization is very much like canonical quantization, where space is continuous, etc. The only difference is because a chosen classical variable has a limited range of values, leading to a focus on a related variable to promote to an operator.

It is generally accepted
      that canonical quantization has not yet produced a satisfactory quantization of Einstein's gravity. Let's really see what affine quantization can do.

     \subsection{Affine quantization and Einstein's gravity}
     In this section we also start with the classical phase-space variables that are used to 
     explore the realm of classical gravity; namely, we again introduce the metric field $g_{ab}(x)$ 
     and the momentum field $\pi^{cd}(x)$ exactly as before. Canonical quantization chooses to promote these two fields to quantum operators, or at least it tries to do that. Affine quantization does not choose these classical variables but replaces the momentum field
     with the affine field $\pi^a_b(x)\equiv \pi^{ac}(x)\,g_{bc}(x)$, with an explicit 
     sum on $c$, and retains the metric field $g_{de}(x)$ along side the affine 
  field.\footnote{This section is partially based on \cite{j1,j2,bqg,mmm}.}
     
     The standard Poisson bracket for the metric and momentum fields is given by 
     \bn \{g_{ab}(x),\pi^{cd}(x')\}=\half\, \de^3(x,x')[\de^c_a\de^d_b+\de^d_a\de^c_b]\;, \en
     and the Poisson brackets for either two metric fields or two momentum fields would vanish. Instead, the set of 
     Poisson brackets for the metric and affine fields is given by
      \bn &&\{\pi^a_b(x),\pi^c_d(x')\}=
   \half\,\delta^3(x,x')\s[\delta^a_d\s \pi^c_b(x)-\delta^c_b\s \pi^a_d(x)\s]\;,    \no \\
       &&\hskip-.20em\{g_{ab}(x), \s \pi^c_d(x')\}= \half\,\delta^3(x,x')\s [\delta^c_a g_{bd}(x)+\delta^c_b g_{ad}(x)\s] \;,      \\
       &&\hskip-.30em\{g_{ab}(x),\s g_{cd}(x')\}=0 \;. \no  \en
       Observe that these Poisson brackets are true even if we change $g_{ab}(x)$ to $-g_{ab}(x)$, and indeed we can even restrict $\{g_{ab}(x)\}>0$. This is not possible with the Poisson bracket for the canonical variables.

         \subsubsection{Affine coherent states for gravity}
    We choose the basic affine operators to build our coherent states for gravity \cite{j1};
    specifcally,
   \bn |\pi;\eta\>=e^{(i/\hbar)\tint \pi^{ab}(x)\hat{g}_{ab}(x)\;d^3\!x} \; e^{-(i/\hbar)\tint\eta^a_b(x)\hat{\pi}^b_a(x)\;d^3\!x}\;|\alpha\>\;\;\;\;[=|\pi;g\>]. \label{drt} \en
   The fiducial vector $|\alpha\>$ has been chosen so that the matrix $\eta(x)\equiv\{\eta^a_b(x)\}$ enters the coherent states solely in the form given by
          \bn \langle\pi;\eta|\hat{g}_{ab}(x)|\pi;\eta\rangle =[e^{\eta(x)/2}]^c_a\,\langle\alpha|\hat{g}_{cd}(x)\,|\alpha\rangle\,[e^{\eta(x)/2}]^d_b\equiv g_{ab}(x)\;, \label{mm} \en
          which preserves metric positivity, i.e., $\{g_{ab}(x)\}>0$.
   A companion relation is given by
    \bn \langle\pi;\eta|\hat{\pi}^a_b(x)|\pi;\eta\rangle
=\pi^{ac}(x)\,g_{cb}(x)\equiv\pi^a_b(x)\;,\label{qq} \en
which involves the metric result from (\ref{mm}). These relations permit us to rename the coherent states from $|\pi;\eta\>$ to $|\pi;g\>$.

               As a consequence, the inner product of two gravity coherent states is given by 
   \bn \langle\pi'';g''|\pi';g'\rangle\hskip-1.3em&&=\exp\Big{\{}\textstyle{-2\int}b(x)\,d^3x \\
   &&\hskip-4em \times\ln\big\{
\frac{\det\{\frac{1}{2}[ {g''}^{ab}(x)+{g'}^{ab}(x)]+i\frac{1}{2\hbar}b(x)^{-1}[{\pi''}^{ab}(x)-{\pi'}^{ab}(x)]\}}{\det[{g''}^{ab}(x)]^{1/2}
\,\,\det[{g'}^{ab}(x)]^{1/2}}  \big\} {\Big\}}\;. \no \en
Here the scalar density function $b(x)>0$ ensures the covariance of this expression.

To test whether or not we have `favored coordinates' we examine, with a suitable factor $J$, the Fubini-Study metric given by
 \bn d\sigma(\pi,g)^2\hskip-1.4em&&\equiv J\hbar[\,\|\,d|\pi;g\>\|^2-|\<\pi;g|\;d|\pi;g>|^2\,] \no \\
     &&=\tint \{(b(x)\hbar)^{-1} g_{ab}(x)\,g_{cd}(x)\,d\pi^{bc}(x)\,d\pi^{da}(x) \\
     &&\hskip2em +(b(x)\hbar) \,g^{ab}(x)\,g^{cd}(x)\,dg_{bc}(x)\,dg_{da}(x)\}\; d^3\!x \;.\no \en
    This metric, like the one in the previous section, represents a multiple family of constant negative curvature spaces. The product of coefficients of the differential terms is proportional to a constant rather like the previous affine metric stories. Based on the previous analysis we accept that
    the basic affine quantum variables have been promoted from basic affine classical variables.
    
    The given choice of coherent states and their quantum operators therein have passed the 
    test to involve constant negative curvature coordinates, which makes them favored affine
    coordinates for an affine quantization.
       
    \subsubsection{Schr\"odinger's representation and equation}
       Passing to operator commutations, the relations (\ref{mm}) and (\ref{qq}) point toward a
 promotion of the set of Poisson brackets to operator commutations given by 
 \bn   &&[\hp^a_b(x),\s \hp^c_d(x')]=i\s\half\,\hbar\,\delta^3(x,x')\s[\delta^a_d\s \hp^c_b(x)-\delta^c_b\s \hp^a_d(x)\s]\;,    \no \\
       &&\hskip-.10em[\hg_{ab}(x), \s \hp^c_d(x')]= i\s\half\,\hbar\,\delta^3(x,x')\s [\delta^c_a \hg_{bd}(x)+\delta^c_b \hg_{ad}(x)\s] \;, \\
       &&\hskip-.20em[\hg_{ab}(x),\s \hg_{cd}(x')] =0 \;. \no  \en
As with the Poisson brackets, these commutators are valid if we change  $\hg_{ab}(x)$ to $-\hg_{ab}(x)$. For the metric and affine fields,
we again find that we can choose the subset for which $\{\hg_{ab}(x)\}>0$. 

The classical Hamiltonian for our models is given \cite{adm} by
        \bn H'(\pi, g)=\tint \{ g(x)^{-1/2} [\pi^a_b(x)\pi^b_a(x)-\half \pi^a_a(x)\pi^b_b(x)] 
           +g(x)^{1/2}\,^{(3)}\!R(x)\}\;d^3x, \label{667} \en
 where $^{(3)}\!R(x)$ is the 3-dimensional Ricci scalar. For the quantum operators
we adopt a Schr\"odinger representation
 for the basic operators: specifically $ \hat{g}_{ab}(x)=g_{ab}(x)$ and 
   \bn \hat{\pi}^a_b(x)=-\half i \hbar\,[\,g_{bc}(x)\,(\delta/\delta\,g_{ac}(x))+(\delta/\delta\,
   g_{ac}(x)))\,g_{bc}(x)\,]\;.\en
 It follows that the Schr\"odinger equation is given by
                \bn && i\hbar\,\d\;\Psi(\{g\}, t)/\d t)=\big\{ \tint \{ [\hat{\pi}^a_b(x)\, g(x)^{-1/2} \,\hat{\pi}^b_a(x)-\half \hat{\pi}^a_a(x)\, g(x)^{-1/2} \,\hat{\pi}^b_b(x)] \no\\
         &&\hskip10em  +g(x)^{1/2}\,^{(3)}\!R(x)\}\;d^3x\;\big\}\;\Psi(\{g\}, t)\;, \label{rrrr} \en
         where $\{g\}$ represents the 
          $\{g_{ab}(x)\}$ matrix field.

     Much like the scalar field of Sec.~2, we expect that the Schr\"odinger representation of eigenfunctions of the Hamiltonian operator have a `large field behavior' and
   a `small field behavior', and the Hamiltonian operator eigenfunctions are formally given by $\Psi(\{g\})=W(\{g\})\,[\Pi_x g(x)^{-1/2}]$, where the `small field behavior' is formally obtained by the relation $ \hp^a_b \s F(g)=0$, which implies that
   $[g_{bc}\s (\d/\d g_{ac})+\half\delta^a_b]\s F(g)=0$ and this leads to $g_{bc}\s g^{ac}\s g\s\s dF(g)/d g + \half\delta^a_b\s F(g)=0$, which requires that
   $ g\s d\s F(g)/d g+\half\s F(g)=0$; hence $F(g)\propto g^{-1/2}$.
   In summary, we observe that
     \bn \hat{\pi}^a_b(x)\,g(x)^{-1/2}=0\;\;\:, \hskip2em \hat{\pi}^a_b(x)\,\Pi_y \,g(y)^{-1/2}=0\;.\label{98} \en
           We next insert a brief, but relative, comment about the Hamiltonian operator constraints.
           
           Using (\ref{98}), 
           the factor $g(x)^{-1/2}$ can be moved to the left in the Hamiltonian density; see
           (\ref{rrrr}). 
           This permits changing the Hamiltonian density, essentially
           by multiplying the Hamiltonian density by $g(x)^{1/2}$, and using that expression to make 
           the result a simpler approach to fulfill the Hamiltonian operator constraints \cite{adm}
           to seek Hilbert space states $\Omega(\{g\})$ such that
           \bn  \big\{ [\hat{\pi}^a_b(x)\,\hat{\pi}^b_a(x)-\half \hat{\pi}^a_a(x) \,
           \hat{\pi}^b_b(x)] 
           +g(x)\,^{(3)}\!R(x)\big\}\;\Omega(\{g\})=0\;.  \label{zero} \en

    As were the procedures in Sec.~2.2, we regularize the chosen eigenfunctions  by replacing the spacial continuum by a set of $N'<\infty$ points labeled by the usual points $\bk a$ and introduce a regularized ($r$) eigenfunction given by
   \bn  \Psi_r(\{g\})=W_r(\{g\})\,\{\Pi_\bk\,(ba^3)^{1/2}\,[\Sigma_\bl J_{\bk,\bl}\,g_\bl]^{-(1-ba^3)/2} \}\;, \en
  where the factors $J_{\bk,\bl}$ are the same factors as in Sec.~2.2. Because the affine variable complex in (\ref{667}) is not positive definite, the quantum eigenvalues will, most likely, range over the whole real line.

    Thus, $W_r(\{g\})$ will, again most likely, be positive and negative for all eigenfunctions, and we focus attention on an
    appropriate eigenfunction that is nonzero in the vicinity of very small values of $g$. Just as in the covariant scalar case, we choose the `large field behavior' of the regularized quantum Hamiltonian operator from the classical Hamiltonian, and we choose the `small field behavior' of the
  regularized quantum Hamiltonian,
   i.e., the term  $\Pi_\bk(ba^3)^{1/2}[\Sigma_\bl J_{\bk,\bl}\,g_\bl]^{-(1-ba^3)/2}$. Based on Sec.~2.4, we are led to the regularized form of the quantum Hamiltonian in the Schr\"odinger density representation given by
      \bn \mfH_r= {\t\sum}_\bk \{ \, {\hp^a_{b\s\bk}\s\s {\bf J}_\bk(g)\s \hp^b_{a\s\bk}} -\half\s{\hp^a_{a\s\bk}\s\s {\bf J}_\bk(g)\s\hp^b_{b\s\bk}}
      +g_\bk^{1/2}\,^{(3)}\!R_\bk  \} \,a^3\;,  \label{eK}\en
      where ${\bf J}_\bk(g)\equiv[\Sigma_\bl J_{\bk,\bl}\,g_\bl]^{-(1-ba^3)/2} $ and
       \bn  &&\hp^a_{b\s \bk}=-i\s\,\half\s\hbar\{\frac{\t\d}{\t\d\s g_{ac\s\bk}}\s\,g_{bc\s\bk}+ g_{bc\s\bk}\,\frac{\t\d}{\t\d\s g_{ac\s\bk}}\} \s a^{-3}\;.\label{eJ}\en

            We have strongly focussed on making the Hamiltonian operator well defined so that, when we
          consider the constraints, we are ensured that the operator will result in the correct properties.


\subsection{Enforcing the constraints}
The classical action functional for gravity is given \cite{adm} by
  \bn A=\tint_0^T\tint\,\{ \pi^{ab}(x,t)\s {\dot g}_{ab}(x,t) - N^a(x,t)\s \pi^b_{a\s |b}(x,t)-N(x,t)\s H(x,t)\,\}\, d^3\!x\,dt\;, \en
  where the Lagrange multipliers, the lapse, $N(x,t)$, and the three shifts, $N^a(x,t)$, enforce the classical Hamiltonian  constraints, $H(x,t)=0$,
  and the classical diffeomorphism constraints, $\pi^b_{a\s |b}(x,t)=0$, for all $x\,\&\, t$. Since the classical constraints are first class, the Lagrange multipliers can assume any
  values in the equations of motion, such as $N(x,t)=1$ and/or $N^a(x,t)$=0. However, in the quantum theory, $H(x,t)$ and $\pi^b_{a\s |b}(x,t)$ become operators, while
  $N(x,t)$ and $N^a(x,t)$ remain classical functions.

  Let us focus on the regularized classical Hamiltonian constraints, $H_\bk=0$, for all $\bk$, and the three regularized classical diffeomorphism
   constraints, $\pi^a_{b\s \bk}$ $\!\! _{|\,a}$ $=0$, for all $b$ and $\bk$,  where $_|$ denotes a regularized
   covariant scalar derivative. The four regularized quantum constraints should follow the classical story as closely as possible, and so, following Dirac,
    we initially propose that vectors in the physical Hilbert space obey
    $\mfH_\bk\s|\Psi\>_{phys}=0$ for all $\bk$ and $\hp^a_{b\s \bk}$ $\!\! _{|\,a}\s|\Psi\>_{phys}=0$ for all $b$ and $\bk$, for a `wide class' of non-zero Hilbert space vectors.
    However, that goal is not possible since, for certain $\bk$ and $\bm$, $[\mfH_\bk,\s\mfH_\bm]\s|\Psi\>_{phys}\ne0$ due to
    quantum second-class constraints. Instead, we choose an appropriate projection operator $\mathbb{E}=\mathbb{E}(N'^{-1}\s[\s\Sigma_\bk\,\mfH^2_\bk+\Sigma_{a,\s\bk}\,\hk^{b\,2}_{a\s\bk|b}\s]\,\le\,\delta(\hbar)^2\,)$, which is adjusted so
    that the constraints have the smallest, non-vanishing values.  If $\<\Psi|\Phi\>$ denotes the inner product in the original, kinematical Hilbert space $\mathcal{H}$, then $\<\Psi|\,\mathbb{E}\,|\Phi\>$ denotes the inner product in the reduced, physical Hilbert space $\mathcal{H}_{phys}$; or symbolically stated, $\mathcal{H}_{phys}=\mathbb{E}\,\mathcal{H}$.

    The projection operator $\mathbb{E}$ can be constructed by a suitable functional integral \cite{proj,ma}.
    In the general case, choosing a set of arbitrary, self-adjoint, constraint operators, $C_\alpha$, where $\alpha\in\{1,2, ..., A\}$, we construct a
    functional integral given by
      \bn\mathbb{E}(\Sigma_\alpha C^2_\alpha\le \delta(\hbar)^2\,)=\int\, \mathbb{T}\, e^{-i\tint_0^T\,\Sigma_\alpha C_\alpha\,\l_\alpha(t)\,dt}\,\mathcal{D} R(\l)\;,\en
      where $\mathbb{T}$ implies a time-ordered integral and $R(\l)$  is a suitable weak measure (see \cite{proj}) which is dependent only on: (i) the time $T>0$, (ii) the upper limit $\delta(\hbar)^2\ge0$, and (iii) the number of constraints $A\le\infty$. {\it The measure $R(\l)$ is completely independent of the choice of the constraint operators $\{C_\alpha\}_{\alpha=1}^N$! }
     
     \subsubsection{A master constraint operator}
      There is an alternative procedure to enforce the quantum constraints as well. Following Thiemann (e.g., \cite{123}), we too can introduce a `Master Constraint Operator' 
to accommodate the Hamiltonian constraints that the Hamiltonian density $H(x)\,
\Omega(\{g\})$ should vanish. Exploiting the relation (\ref{zero}), we introduce
  \bn {\cal{M}}\equiv \tint \big\{ [\hat{\pi}^a_b(x)\,\hat{\pi}^b_a(x)-\half \hat{\pi}^a_a(x) \,
           \hat{\pi}^b_b(x)] 
           +g(x)\,^{(3)}\!R(x)\big\}\;g(x)^{-2}\; \no \\
           \times\big\{ [\hat{\pi}^a_b(x)\,\hat{\pi}^b_a(x)-\half 
           \hat{\pi}^a_a(x) \,
           \hat{\pi}^b_b(x)] 
           +g(x)\,^{(3)}\!R(x)\big\}\;g(x)^{1/2}\;d^3\!x \;, \en
           and thus $ {\cal{M}}\;\Omega(\{g\})=0$ for all vectors in the physical Hilbert space.
          Indeed, exploiting (\ref{98}), we can simplify the last equation to become
          \bn {\cal{M}}=\tint\big\{ [\hat{\pi}^a_b(x)\,\hat{\pi}^b_a(x)-\half \hat{\pi}^a_a(x) \,
           \hat{\pi}^b_b(x)]
           +g(x)\,^{(3)}\!R(x)\big\}^2\;g(x)^{-3/2}\;d^3\!x \;.\en
           
           The other constraints for gravity are the three equations $\pi^b_{a|b}(x)=0$. 
       We can deal with these constraints by constructing
           \bn {\cal{N'}}\equiv \tint\{ \pi^b_{a|b}(x)\,g^{ac}(x)\,\pi^d_{c|d}(x)\}\;g(x)^{_-1/2}\;d^3\!x\;.\en
           Finally, we can include all constraints in
           \bn {\cal{L}}\equiv {\cal{M}}+{\cal{N}}'\;. \en
           Physical Hilbert states $\Omega(\{g\})$ are those for which ${\cal{L}}\;\Omega(\{g\})=0$, while $\Omega(\{g\})\neq0$. 
           
           To offer an example of a few vectors that are in the physical Hilbert space, it helps to reduce the underlying spacial space to a finite level. In that case,
           the vector $\Omega(\{g\})= g(x)^{-1/2}$ for which $g(x)=\det(g_{ab}(x))$, where, e.g.
           $g_{11}(x)=3.2$, $g_{22}(x)=1.7$, $g_{33}(x)=2.4$, and $g_{12}(x)=g_{21}(x)=0.34$; all 
           other 
           elements are zero. Let us call this particular example  $\Omega_1(\{g\})$, namely the first 
           example. 
           A second example is $\Omega_2(\{g\})$, with a different set of constant values, and that
           type of vector can also lead to $\Omega_a(\{g\})=0.8\,\Omega_1(\{g\})+1.2(1+i)\,\Omega_2(\{g\})$,
           etc. 
           
           Admittedly, these are simple vectors, but nevertheless, they are vectors in the physical 
           Hilbert space. Clearly, more vectors are needed.

          \section*{Conclusion}
   If the reader can accept that an `harmonic 
   oscillator' for which $0<q<\infty$ can not be quantized by canonical quantization but can be quantized by affine quantization (which is
  demonstrated in \cite{j1}), then it is a natural step to examine the affine quantization of  non-renormalizable scalar fields and Einstein's gravity, with both not having been generally accepted as being  successfully quantized by canonical quantization. Affine quantization used for these same problems offers entirely reasonable solutions, despite their complex results. 
  
  For many years the author has 
  recognized the possibilities of affine quantization, which imitate all of the procedures of canonical quantization, but differs only by a
  different pair of basic quantum operators that also have their roots in appropriate classical theories; a focussed lesson regarding affine quantization appears in \cite{j2}. 
  Perhaps there are other areas of theoretical physics that could profit from exploiting the power of affine quantization.

\end{document}